\documentclass[twocolumn,showpacs,preprintnumbers]{revtex4}%
\usepackage{amssymb}
\usepackage{amsmath}
\usepackage{graphicx}
\usepackage{dcolumn}
\usepackage{bm}
\usepackage{amsfonts}%
\begin{document}
\title{Quantum Nondemolition Measurement of Discrete Fock States of a Nanomechanical Resonator}
\author{Eyal Buks, Eran Segev, Stav Zaitsev, Baleegh Abdo}
\affiliation{Department of Electrical Engineering, Technion, Haifa 32000 Israel}
\author{M. P. Blencowe}
\affiliation{Department of Physics and Astronomy, Dartmouth College, Hanover, New Hampshire
03755, USA}
\date{\today }

\begin{abstract}
We study theoretically a radio frequency superconducting interference device
integrated with a nanomechanical resonator and an LC resonator. By applying
adiabatic and rotating wave approximations, we obtain an effective Hamiltonian
that governs the dynamics of the mechanical and LC resonators. Nonlinear terms
in this Hamiltonian can be exploited for performing a quantum nondemolition
measurement of Fock states of the nanomechanical resonator. We address the
feasibility of experimental implementation and show that the nonlinear
coupling can be made sufficiently strong to allow the detection of discrete
mechanical Fock states.

\end{abstract}
\maketitle





Anharmonicity may introduce coupling between different modes of a resonator.
In a seminal paper \cite{Sanders_694}, Sanders and Milburn found that
inter-mode coupling could in principle enable quantum non-demolition (QND)
detection \cite{Braginsky_QM,Caves_341} of discrete Fock states of a
\textit{signal} mode by intensively driving another nonlinearly coupled
\textit{detector} (or pump) mode, and monitoring the response near the pump
frequency. Such a measurement scheme is characterized by a measurement time
$\tau_{m}$, defined as the time needed to distinguish between initial states
of the signal mode having different Fock numbers. Fock number detection can be
realized if $\tau_{m}$ can be made shorter than the lifetime of a Fock state
of the signal mode.

The prospects of employing this technique for QND measurement of a single
phonon in a mesoscopic mechanical resonator was recently studied theoretically
in Ref. \cite{Santamore_144301}. In practice, however, anharmonic couplings
between different modes of a mechanical resonator are typically far too weak
to allow single-phonon detection when the detector mode is taken to have a
linear response. On the other hand, a significant enhancement can be achieved
by driving the detector mode into the nonlinear regime and exploiting critical
slowing down by operating close to the edge of the region where the response
exhibits bistability \cite{Santamore_052105,Buke_023815}.%

\begin{figure}
[ptb]
\begin{center}
\includegraphics[
height=1.6336in,
width=2.5399in
]%
{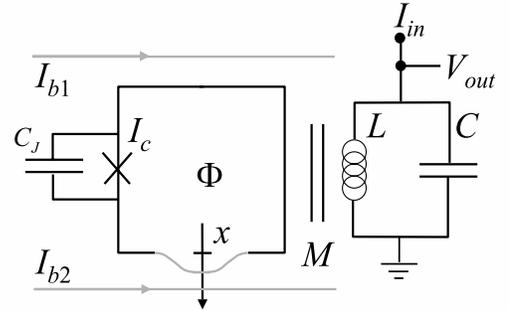}%
\caption{The device consists of an RF SQUID integrated with a nanomechanical
and an LC resonators. The external flux is applied by using 2 bias current
lines carrying dc currents $I_{b1}$ and $I_{b2}$ respectively. One of the
current line lies very close to the vibrating beam, allowing thus a relatively
high value of the magnetic field $B$ at the location of the beam. On the other
hand, the Josephson junction is located symmetrically in such a way that the
magnetic field at its location vanishes when $I_{b1}=I_{b2}$. This
configuration allows applying the needed external flux to the loop $\Phi
_{e}=\Phi_{0}/2$ without significantly degrading the critical current of the
junction provided that the area of the junction is kept much smaller than the
area of the loop.}%
\label{device}%
\end{center}
\end{figure}

To achieve single phonon sensitivity, it is highly desirable to increase the
anharmonic coupling between the signal mode and the detector mode. Here we
propose a novel configuration in which a fundamental mode of a doubly clamped
beam, serving as the mechanical signal mode, is indirectly coupled to an LC
resonator serving as the detector. The anharmonic coupling is achieved by
coupling both the mechanical mode and the LC resonator directly to a radio
frequency superconducting interference device (RF SQUID) \cite{Buks_174504}.
We show that the RF SQUID degree of freedom can be eliminated from the
equations of motion by employing an adiabatic approximation. Moreover,
employing the rotating wave approximation (RWA) leads to further
simplification of the equations of motion of the system. We find that the
effective anharmonic coupling constant between the mechanical mode and the LC
resonator in this configuration can be made sufficiently large to allow single
phonon detection under appropriate conditions.

Alternative approaches for performing QND measurement of the energy of a
nanomechanical resonator were proposed recently in Refs.
\cite{Jacobs_147201,Jacobs_0707_3803}. Similar systems consisting of a SQUID
integrated with a nanomechanical resonator have been recently studied
theoretically \cite{Blencowe_014511,Buks_0705_0206}. Zhou and Mizel have shown
that nonlinear coupling between a DC SQUID and a mechanical resonator can be
employed for producing squeezed states of the mechanical resonator
\cite{Zhou_0605017}. More recently, Xue \textit{et al}. have shown that a flux
qubit integrated with a nanomechanical resonator can form a cavity quantum
electrodynamics system in the strong coupling region \cite{Xue_35}.

The device, which is seen in Fig. \ref{device}, consists of an RF SQUID
inductively coupled to an LC resonator. A section of the RF SQUID loop is
freely suspended and allowed to oscillate mechanically. We assume the case
where the fundamental mechanical mode vibrates in the plane of the loop and
denote the amplitude of this flexural mode as $x$. Let $m$ be the effective
mass of the fundamental mode, and $\omega_{m}$ its angular resonance
frequency. A magnetic field is applied perpendicularly to the plane of the
loop. Let $\Phi_{e}$ be the externally applied flux for the case $x=0$, and
let $B$ denote the component of the magnetic field normal to the plane of the
loop at the location of the doubly clamped beam (it is assumed that $B$ is
constant in the region where the beam oscillates). A Josephson junction (JJ)
having a critical current $I_{c}$ and capacitance $C_{J}$ is integrated into
the loop. The RF SQUID is inductively coupled to a resonator comprising an
inductor $L$ and a capacitor $C$ in parallel. The mutual inductance between
the RF SQUID and the resonator is $M$. Detection is performed by injecting a
monochromatic input current $I_{in}$ into the LC resonator and measuring the
output voltage $V_{out}$ (see Fig. \ref{device}).

The total magnetic flux $\Phi$ threading the loop is given by $\Phi=\Phi
_{e}+Blx+\Phi_{i}$, where $l$ is an effective length of the beam. The term
$\Phi_{i}=I_{s}\Lambda+MI_{L}$ represents the flux generated by both the
circulating current in the RF SQUID $I_{s}$ and by the current in the inductor
of the LC resonator $I_{L},$where $\Lambda$ is the self inductance of the
loop. Similarly, the magnetic flux in the inductor of the LC resonator is
given by $\varphi=I_{L}L+MI_{s}$.

The gauge invariant phase across the Josephson junction is given by
$\theta=2\pi n-2\pi\Phi/\Phi_{0}$, where $n$ is integer and $\Phi_{0}=h/2e$ is
the flux quantum.

The Lagrangian of the closed system is expressed as a function of $x$,
$\varphi$ and $\Phi$ and their time derivatives (denoted by overdot)%

\begin{equation}
\mathcal{L}=\frac{m\dot{x}^{2}}{2}+\frac{C\dot{\varphi}^{2}}{2}+\frac
{C_{J}\dot{\Phi}^{2}}{2}-U_{0}-U_{1}\ ,
\end{equation}
where the potential terms are given by%

\begin{subequations}
\begin{align}
U_{0}  &  =\frac{m\omega_{m}^{2}x^{2}}{2}+\frac{C\omega_{e}^{2}\varphi^{2}}%
{2}-I_{in}\varphi\ ,\\
U_{1}  &  =\frac{\left(  \Phi-\Phi_{e}-Blx-\frac{M\varphi}{L}\right)  ^{2}%
}{2\Lambda\left(  1-K^{2}\right)  }-\frac{\Phi_{0}I_{c}\cos\frac{2\pi\Phi
}{\Phi_{0}}}{2\pi}\ ,
\end{align}
and where $\omega_{e}=1/\sqrt{LC}$ and $K=M/\sqrt{\Lambda L}$.

The Euler - Lagrange equations can be written as%
\end{subequations}
\begin{subequations}
\begin{align}
m\ddot{x}+m\omega_{m}^{2}x-BlI_{s}  &  =0\ ,\label{x dot dot}\\
C\ddot{\varphi}+I_{L}-I_{in}  &  =0\ ,\label{varphi dot dot}\\
C_{J}\ddot{\Phi}+I_{s}+I_{c}\sin\frac{2\pi\Phi}{\Phi_{0}}  &  =0\ .
\label{Phi dot dot}%
\end{align}
The interpretation of these equations is straightforward. Eq. (\ref{x dot dot}%
) is Newton's 2nd law for the mechanical resonator, where the force is
composed of the restoring elastic force $-m\omega_{m}^{2}x$ and the Lorentz
force $BlI_{s}$ acting on the movable beam. Eq. (\ref{varphi dot dot}) states
that the injected current $I_{in}$ into the LC resonator equals the sum of the
current in the inductor $I_{L}$ and the one in the capacitor $C\ddot{\varphi}%
$. Similarly, Eq. (\ref{Phi dot dot}) states that the circulating current
$I_{s}$ equals the sum of the current $I_{c}\sin\theta$ through the JJ and the
current $C_{J}\dot{V}_{J}$ through the capacitor, where the voltage $V_{J}$
across the JJ is given by the second Josephson equation $V_{J}=\left(
\Phi_{0}/2\pi\right)  \dot{\theta}$.

The variables canonically conjugate to $x$, $\varphi$ and $\Phi$ are given by
$p=\partial\mathcal{L}/\partial\dot{x}$, $q=\partial\mathcal{L}/\partial
\dot{\varphi}$ and $Q=\partial\mathcal{L}/\partial\dot{\Phi}$ respectively.
The Hamiltonian is given by
\end{subequations}
\begin{equation}
\mathcal{H}=\mathcal{H}_{0}+\mathcal{H}_{1}\ ,
\end{equation}
where
\begin{subequations}
\begin{align}
\mathcal{H}_{0} &  =\frac{p^{2}}{2m}+\frac{q^{2}}{2C}+U_{0}\ ,\\
\mathcal{H}_{1} &  =\frac{Q^{2}}{2C_{J}}+U_{1}\ .
\end{align}
Quantization is achieved by regarding the variables $x,$ $p,$ $\varphi,$ $q,$
$\Phi$ and $Q$ as Hermitian operators satisfying canonical commutation relations.

As a basis for expanding the general solution we use the eigenvectors of the
following Schr\"{o}dinger equation%
\end{subequations}
\begin{equation}
\mathcal{H}_{1}\left\vert n\left(  x,\varphi\right)  \right\rangle
=\varepsilon_{n}\left(  x,\varphi\right)  \left\vert n\left(  x,\varphi
\right)  \right\rangle \ , \label{e.v H_1}%
\end{equation}
where $x$ and $\varphi$ are treated here as parameters (rather than degrees of
freedom). The local eigen-vectors are assumed to be orthonormal $\left\langle
m\left(  x,\varphi\right)  |n\left(  x,\varphi\right)  \right\rangle
=\delta_{nm}$.

The eigenenergies $\varepsilon_{n}\left(  x,\varphi\right)  $ and the
associated wavefunctions $\vartheta_{n}$ are found by solving the following
Schr\"{o}dinger equation%
\begin{equation}
\left(  -\beta_{C}\frac{\partial^{2}}{\partial\phi^{2}}+u\right)
\vartheta_{n}=\frac{\varepsilon_{n}}{E_{0}}\vartheta_{n}\ .
\label{Scrodinger phi}%
\end{equation}
where%

\begin{align}
u  &  =\frac{\left(  \phi-\phi_{0}\right)  ^{2}}{1-K^{2}}+2\beta_{L}\cos
\phi\ ,\label{u}\\
\phi &  =\frac{2\pi}{\Phi_{0}}\left(  \Phi-\frac{\Phi_{0}}{2}\right)  \ ,\\
\phi_{0}  &  =\frac{2\pi}{\Phi_{0}}\left(  \Phi_{e}-\frac{\Phi_{0}}%
{2}+Blx+\frac{M\varphi}{L}\right)  \ ,
\end{align}
$\beta_{L}=2\pi\Lambda I_{c}/\Phi_{0}$, $\beta_{C}=2e^{2}/C_{J}E_{0}$ and
$E_{0}=\Phi_{0}^{2}/8\pi^{2}\Lambda$.

The total wave function is expanded as
\[
\psi=\sum\nolimits_{n}\xi_{n}\left(  x,\varphi,t\right)  \left\vert
n\right\rangle \ .
\]
In the adiabatic approximation \cite{Moody_160} the time evolution of the
coefficients $\xi_{n}$ is governed by the following set of decoupled equations
of motion%
\begin{equation}
\left[  \mathcal{H}_{0}+\varepsilon_{n}\left(  x,\varphi\right)  \right]
\xi_{n}=i\hbar\dot{\xi}_{n}\ .
\end{equation}
Note that in the present case the geometrical vector potential
\cite{Moody_160} vanishes. The validity of the adiabatic approximation will be
discussed below.

In what follows we focus on the case where $\left\vert \phi_{0}\right\vert
\ll1$ and $\beta_{L}\left(  1-K^{2}\right)  >1$. In this case the adiabatic
potential $u\left(  \phi\right)  $ given by Eq. (\ref{u}) contains two wells
separated by a barrier near $\phi=0$. At low temperatures only the two lowest
energy levels contribute. In this limit the local Hamiltonian $\mathcal{H}%
_{1}$ can be expressed in the basis of the states $\left\vert \curvearrowleft
\right\rangle $ and $\left\vert \curvearrowright\right\rangle $, representing
localized states in the left and right well respectively having opposite
circulating currents. In this basis, $\mathcal{H}_{1}$ is represented by the
$2\times2$ matrix%
\begin{equation}
\mathcal{H}_{1}\dot{=}\left(
\begin{array}
[c]{cc}%
\eta\phi_{0} & \Delta\\
\Delta & -\eta\phi_{0}%
\end{array}
\right)  \ . \label{2-level H_1}%
\end{equation}
The real parameters $\eta$ and $\Delta$ can be determined by solving
numerically the Schr\"{o}dinger equation (\ref{Scrodinger phi}). The
eigenenergies are given by%
\begin{equation}
\varepsilon_{\pm}=\pm\sqrt{\eta^{2}\phi_{0}^{2}+\Delta^{2}}\ .
\end{equation}

It is convenient to introduce the annihilation operators%
\begin{align}
A_{m} &  =\frac{e^{i\omega_{m}t}}{\sqrt{2\hbar}}\left(  \sqrt{m\omega_{m}%
}x+\frac{i}{\sqrt{m\omega_{m}}}p\right)  \ ,\\
A_{e} &  =\frac{e^{i\omega_{e}t}}{\sqrt{2\hbar}}\left(  \sqrt{C\omega_{e}%
}\varphi+\frac{i}{\sqrt{C\omega_{e}}}q\right)  \ ,
\end{align}
and the corresponding number operators $N_{m}=A_{m}^{\dag}A_{m}$ and
$N_{e}=A_{e}^{\dag}A_{e}$. Consider the case where $\Phi_{e}=\Phi_{0}/2$, and
assume that adiabaticity holds and the RF SQUID remains in its lowest energy
state. The energy of this state $\varepsilon_{-}\left(  x,\varphi\right)  $
can be expressed in terms of the annihilation operators $A_{m}$, $A_{e}$ and
their Hermitian conjugates. The resulting expression generally contains terms
oscillating at frequencies $\omega_{k,l}=k\omega_{m}+l\omega_{e}$, where $k$
and $l$ are integers. In the RWA such terms are neglected unless $\omega
_{k,l}=0$ or $\omega_{k,l}$ is close to the frequency of the externally
injected bias current $I_{in}$, since otherwise the effect on the dynamics on
a time scale much longer than a typical oscillation period is negligibly
small. Keeping terms up to fourth order in $\phi_{0}$ the eigenenergy
$\varepsilon_{-}$ in the RWA can be expressed as%
\begin{align}
\varepsilon_{\mathrm{RWA}}\left(  x,\varphi\right)   &  =\hbar(\Omega
_{0}+\Omega_{2,0}N_{m}+\Omega_{0,2}N_{e}\nonumber\\
&  +\Omega_{4,0}N_{m}^{2}+\Omega_{0,4}N_{e}^{2}+\Omega_{2,2}N_{m}%
N_{e})\ .\nonumber\\
&
\end{align}
The constant term $\hbar\Omega_{0}$ can be disregarded, since it only gives
rise to a constant phase factor. Moreover, the linear terms $\hbar\Omega
_{2,0}N_{m}$ and $\hbar\Omega_{0,2}N_{e}$ only give rise to a small
renormalization of the resonance frequencies $\omega_{m}$ and $\omega_{e}$
respectively. In addition, since no external drive is applied directly to the
mechanical oscillator, the expectation value $\left\langle N_{m}\right\rangle
$ is expected to be relatively small and consequently the term proportional to
$N_{m}^{2}$ can be neglected. Thus, in the RWA the Hamiltonian $\mathcal{H}%
_{0}+\varepsilon_{-}$, which governs the dynamics of the mechanical and LC
resonators, is given by
\begin{align}
\mathcal{H}_{\mathrm{RWA}} &  =\hbar\omega_{m}N_{m}+\hbar\omega_{e}%
N_{e}\nonumber\\
&  -I_{in}\sqrt{\frac{\hbar}{2C\omega_{e}}}\left(  e^{-i\omega_{e}t}%
A_{e}+e^{i\omega_{e}t}A_{e}^{\dag}\right)  \nonumber\\
&  +\hbar\Omega_{0,4}N_{e}^{2}+\hbar\Omega_{2,2}N_{m}N_{e}\ ,\nonumber\\
&
\end{align}
where%
\begin{equation}
\Omega_{0,4}=\frac{3\Delta}{4\hbar}\left(  \frac{\eta}{\Delta}\frac{2\pi
M}{\Phi_{0}L}\sqrt{\frac{\hbar}{2C\omega_{e}}}\right)  ^{4}\ ,
\end{equation}%
\begin{equation}
\Omega_{2,2}=\frac{3\Delta}{\hbar}\left(  \frac{\eta}{\Delta}\frac{2\pi
Bl}{\Phi_{0}}\sqrt{\frac{\hbar}{2m\omega_{m}}}\right)  ^{2}\left(  \frac{\eta
}{\Delta}\frac{2\pi M}{\Phi_{0}L}\sqrt{\frac{\hbar}{2C\omega_{e}}}\right)
^{2}\ .
\end{equation}

All subsequent analysis is based on Refs.
\cite{Sanders_694,Santamore_144301,Santamore_052105,Buke_023815}, which
consider how nonlinear terms of the kind appearing in the Hamiltonian
$\mathcal{H}_{\mathrm{RWA}}$ can be utilized for QND detection of discrete
Fock states. A single phonon added to the mechanical resonator results in a
shift in the effective resonance frequency of the LC resonator. Consider first
the case $\Omega_{0,4}=0$, where the response of the LC resonator is linear,
and thus this frequency shift is given by $\Omega_{2,2}$. The effective
resonance frequency can be continuously monitored by employing an homodyne
detection scheme, in which $V_{out}$ is mixed with a local oscillator at the
same frequency as the frequency of the driving current $I_{in}$. The
measurement time $\tau_{m}$, which is needed to detect a frequency shift
$\Omega_{2,2}$, is given in the high temperature limit $k_{B}T\gg\hbar
\omega_{e}$ by \cite{Santamore_144301,Buks_0606081,Buks_0705_0206}%
\begin{equation}
\tau_{m}=\frac{2\pi\gamma_{e}}{\Omega_{2,2}^{2}}\frac{k_{B}T}{U_{0}}\ ,
\end{equation}
where $U_{0}=\left\langle N_{e}\right\rangle \hbar\omega_{e}$ is the stored
energy in the stripline resonator. Single phonon sensitivity is achieved when
$\zeta\gtrsim1$, where the dimensionless parameter $\zeta$ is defined as
$\zeta\equiv t_{0}/\tau_{m}$, and $t_{0}$ is a characteristic lifetime of a
Fock state of the mechanical resonator. At a high temperature $k_{B}T\gg
\hbar\omega_{m}$ the lifetime is given by $t_{0}^{-1}\simeq\gamma_{m}\left(
k_{B}T/\hbar\omega_{m}\right)  ^{2}$, where $\gamma_{m}$ is the damping rate
of the mechanical mode and $k_{B}T$ is the thermal energy (see Eq. (55) of
Ref. \cite{Santamore_144301}).

When, however, $\Omega_{0,4}\neq0$, the response of the LC resonator is
approximately linear only when $\left\langle N_{e}\right\rangle $ is smaller
than the critical value corresponding to the onset of nonlinear bistability.
Using this critical value $\left\langle N_{e}\right\rangle _{c}=\gamma
_{e}/\sqrt{3}\Omega_{0,4}$, which is given by Eq. (38) of Ref.
\cite{Buke_023815}, one finds that the largest possible value of $\zeta$ in
the linear regime is roughly given by%
\begin{align}
\zeta_{\mathrm{\max}}  &  \simeq\frac{12\Delta}{\sqrt{3}\hbar\gamma_{m}%
}\left(  \frac{\eta}{\Delta}\frac{2\pi Bl}{\Phi_{0}}\sqrt{\frac{\hbar
}{2m\omega_{m}}}\right)  ^{4}\left(  \frac{\hbar\omega_{m}}{k_{B}T}\right)
^{2}\frac{\hbar\omega_{e}}{k_{B}T}\nonumber\\
&  =\frac{2.8\times10^{-15}\times\frac{\Delta}{h%
\operatorname{GHz}%
}\frac{\omega_{m}}{\gamma_{m}}\left(  \frac{\eta}{\Delta}\frac{B}{%
\operatorname{T}%
}\frac{l}{%
\operatorname{\mu m}%
}\right)  ^{4}\frac{\omega_{e}}{2\pi%
\operatorname{GHz}%
}}{\frac{\omega_{m}}{2\pi%
\operatorname{GHz}%
}\left(  \frac{m}{10^{-19}%
\operatorname{kg}%
}\right)  ^{2}\left(  \frac{T}{%
\operatorname{K}%
}\right)  ^{3}}\ .\nonumber\\
&  \label{zeta_max}%
\end{align}
As was mentioned above, operating the LC resonator pump mode in the regime of
nonlinear response may allow a significant enhancement in the sensitivity by
driving the LC resonator to a jump point at the edge of the region of
bistability \cite{Santamore_144301,Buke_023815,Buks_0705_0206}. Note, however,
that the perturbative approach employed in Ref. \cite{Buke_023815} is not
valid close to the edge of the region of bistability, and thus further study
is required to analyze the behavior of the system in this region.

We now return to the adiabatic approximation and examine its validity. As
before, consider the case where the externally applied flux is given by
$\Phi_{e}=\Phi_{0}/2$, and the LC resonator is driven close to the onset of
nonlinear bistability where the number of photons approaches the critical
value $\left\langle N_{e}\right\rangle _{c}$. Using Refs.
\cite{Buks_174504,Buks_628} one finds that adiabaticity holds, namely Zener
transitions between adiabatic states are unlikely, provided that%
\begin{equation}
\frac{\pi\Delta^{2}}{\eta\hbar\Gamma_{c}}\gtrsim1\ ,
\end{equation}
where%
\begin{equation}
\Gamma_{c}=\frac{2\pi M}{\Phi_{0}L}\sqrt{\frac{2\left\langle N_{e}%
\right\rangle _{c}\hbar\omega_{e}}{C}}\ .
\end{equation}

As can be seen by examining Eq. (\ref{zeta_max}), satisfying the condition
$\zeta_{\mathrm{\max}}\gtrsim1$ together with all other requirements is quite
difficult. However, as we show below, a careful design together with taking
full advantage of recent technological progress may open the way towards
experimental implementation. We discuss below some key considerations. Due to
the relatively strong dependence of $\zeta_{\mathrm{\max}}$ on $\eta/\Delta$
and on $Bl$, it is important to minimize the uncertainty in the values of
these parameters to allow a proper design. The parameter $Bl$ plays a crucial
role in determining the coupling strength between the mechanical resonator and
the RF SQUID. Enhancing the coupling can be achieved by increasing the applied
magnetic field at the location of the mechanical resonator $B$. However, $B$
should not exceed the superconducting critical field. Moreover, the externally
applied magnetic field at the location of the JJ must be kept at a much lower
value in order to minimize an undesirable reduction in $I_{c}$. This can be
achieved by employing an appropriate design in which the applied field is
strongly nonuniform (see Fig. \ref{device}). In addition the sensitivity of
the device can be further enhanced by increasing the ratio $\eta/\Delta$. To
increase the value of $\eta$ it is desirable to employ a JJ having a high
plasma frequency. On the other hand, the value of the energy gap $\Delta$ can
be reduced by increasing $\beta_{L}$. Note, however, that $\Delta$ has to be
kept larger than $k_{B}T$ to ensure that thermal population of the first
excited state of the RF SQUID is negligible. In addition, a successful
experimental implementation requires a very low temperature $T$, low mass $m$,
high resonance frequency $\omega_{e}$, and low mechanical damping $\gamma_{m}%
$. Note also that, as was pointed out before, an additional enhancement of
$\zeta$ (beyond the linear value $\zeta_{\mathrm{\max}}$) may be achieved by
exploiting nonlinearity \cite{Santamore_144301,Buke_023815,Buks_0705_0206}.

As an example, we consider below a device in which a single-walled carbon
nanotube, which serves as the nanomechanical resonator
\cite{Sazonova_284,Witkamp_2904,Peng_087203}, is integrated into the loop of
an RF SQUID
\cite{Cleuziou_53,Bezryadin_971,Kasumov_214521,Friedman_43,Koch_1216}. To
enhance the plasma frequency we assume the case where a microbridge weak link
is employed as a Josephson junction in the RF SQUID \cite{Ralph_10753}. The
following parameters are assumed: $C=3.2\times10^{-11}%
\operatorname{F}%
$, $\omega_{e}/2\pi=0.22%
\operatorname{GHz}%
$ and $\omega_{e}/\gamma_{e}=10^{4}$ for the LC resonator, $\Lambda
=1.1\times10^{-10}%
\operatorname{H}%
$, $C_{J}=10^{-16}%
\operatorname{F}%
$ and $I_{c}=12%
\operatorname{\mu A}%
$ for the RF SQUID, $m=10^{-19}%
\operatorname{kg}%
$, $\omega_{m}/2\pi=0.5%
\operatorname{GHz}%
$ and $\omega_{m}/\gamma_{m}=10^{3}$ for the mechanical resonator, temperature
$T=0.02%
\operatorname{K}%
$, and $K=0.001$ and $Bl=0.05%
\operatorname{T}%
\times%
\operatorname{\mu m}%
$ for the coupling between the RF SQUID and between the LC and mechanical
resonators respectively. These example parameters are achievable with present
day technology. The chosen value of $\Lambda$ corresponds to a circular loop
with a radius of about $18%
\operatorname{\mu m}%
$ and a wire having a cross section radius of about $0.22%
\operatorname{\mu m}%
$, whereas the values of $C_{J}$ and $I_{c}$ correspond to a junction having a
plasma frequency of about $1.2%
\operatorname{THz}%
$. Using these values one finds $\beta_{L}=3.9$ and $\beta_{C}=1.0$. The
values of $\beta_{L}$, $\beta_{C}$ and $K$ are employed for calculating
numerically the eigenstates of Eq. (\ref{Scrodinger phi}) \cite{Buks_174504}.
From these results one finds for the values of the $\eta$ and $\Delta$
parameters in the two-level approximation to Hamiltonian $\mathcal{H}_{1}$
[Eq. (\ref{2-level H_1})] $\eta=4.8E_{0}$ and $\Delta=1.2\times10^{-3}E_{0}$.
Using these values yields $\Omega_{2,2}/2\pi=220%
\operatorname{kHz}%
$, $\Delta/h=0.92%
\operatorname{GHz}%
$,%
\begin{subequations}
\begin{align}
\zeta_{\mathrm{\max}}  &  =230\ ,\label{xi_max nu}\\
\frac{\pi\Delta^{2}}{\eta\hbar\Gamma_{c}}  &  =170\ . \label{-log(p) nu}%
\end{align}
Eq. (\ref{xi_max nu}) indicates that single phonon detection is feasible, even
when the response of the LC resonator is assumed to be linear. Moreover, Eq.
(\ref{-log(p) nu}) ensures the validity of the adiabatic approximation.

In summary, we have demonstrated that an RF SQUID can be exploited for
introducing a tailored coupling between a mechanical and LC resonator, and
that such coupling can be made sufficiently strong to allow a QND measurement
of discrete mechanical Fock states. According to the Bohr's complementarity
principle \cite{Bohr_200}, in the limit of single-phonon sensitivity,
dephasing is expected to come into play, leading to broadening of the
resonance line shape of the mechanical signal mode, allowing thus a relatively
simple experimental verification by performing a \textit{which-path} like experiment.

We than Yuval Yaish for a valuable discussion. This work is partly supported
by the US - Israel Binational Science Foundation (BSF), Israel Science
Foundation, and by the Israeli Ministry of Science.

\bibliographystyle{apsrev}
\bibliography{acompat,Eyal_Bib}

\end{subequations}
\end{document}